\documentclass[12pt]{iopart}
\usepackage[T1]{fontenc}
\usepackage[latin1]{inputenc}
\usepackage[english]{babel}
\usepackage{iopams}

\usepackage[usenames,dvipsnames]{pstricks}
\usepackage{epsfig}

\begin{document}

\title[Universal scaling for second class particles]{Universal scaling for second class particles in a one-dimensional misanthrope process}
\author{Attila R\'akos}
\address{Research Group for Condensed Matter Physics of the Hungarian Academy of Sciences, Budafoki u.\ 8, 1111 Budapest, Hungary}
\ead{rakos@phy.bme.hu}

\begin{abstract}
We consider the one-dimensional Katz-Lebowitz-Spohn (KLS) model, which is a two-parameter generalization of the Totally Asymmetric Simple Exclusion Process (TASEP) with nearest neighbour interaction. Using a powerful mapping, the KLS model can be translated into a misanthrope process. In this model, for the repulsive case, it is possible to introduce second class particles, the number of which is conserved. We study the distance distribution of second class particles in this model numerically and find that for large distances it decreases as $x^{-3/2}$. This agrees with a previous analytical result for the TASEP where the same asymptotic behaviour was found \cite{Derrida1993c}. We also study the dynamical scaling function of the distance distribution and find that it is universal within this family of models.
\end{abstract}


\section{Introduction}

The study of stochastic particle systems with second class particles (SCP) \cite{Ferrari1991} is twofold. On the one hand these systems provide relatively simple examples for particle systems with two conserved densities \cite{Schutz2003}, since the dynamics of first class particles are independent of SCPs, while SCPs move stochastically on a landscape determined by the first class particles. 
On the other hand, the introduction of second class particles is often a powerful tool for studying fluctuations in the original model (with only first class particles) \cite{Beijeren1993,Ferrari1994,Balazs2007}, because SCPs in general follow the trajectories of density perturbations of first class particles. About the relationship between current fluctuations and fluctuations in the position of SCPs see e.g. \cite{Balazs2007} and \cite{Beijeren1993}. Since SCPs are in general attracted by shocks which often appear in one-dimensional driven diffusive systems, SCPs can also be used for a good definition of the microscopic position of shocks \cite{Derrida1993c,Ferrari1991,Andjel1988,Ferrari1992,Derrida1997a,Balazs2001,Balazs2010}.

In the present work we study the effective interaction between SCPs which is mediated by the common density landscape of first class particles. An important, and probably the most widely known example of a system where second class particles can be introduced is the one-dimensional Asymmetric Simple Exclusion Process (ASEP). The ASEP is a lattice model where each lattice site is either occupied by a particle or empty, double occupancy is forbidden. Particles can hop to the left and right nearest neighbour sites with rates $p$ and $q$ ($p\neq q$) if they are empty \cite{Liggett1999,Schutz2001}. In this model second class particles can be introduced by the following dynamical rules \cite{Derrida1993c,Ferrari1991,Schutz2001}:
\begin{equation}
\label{ASEP_SCP}
\begin{array}{ll}
 1 0 \rightarrow 0 1 & \mbox{with rate } p \\ 
 0 1 \rightarrow 1 0 & \mbox{with rate } q \\ 
 2 0 \rightarrow 0 2 & \mbox{with rate } p \\ 
 0 2 \rightarrow 2 0 & \mbox{with rate } q \\ 
 1 2 \rightarrow 2 1 & \mbox{with rate } p \\ 
 2 1 \rightarrow 1 2 & \mbox{with rate } q
\end{array}
\end{equation}
Here $1$ and $2$ indicates first and second class particles respectively, while $0$ is a vacancy. Notice that by not distinguishing between first and second class particles the dynamics in (\ref{ASEP_SCP}) reduces to that of the ASEP without SCPs. The same is the result if SCPs are not distinguished from vacancies. 

A breakthrough in the analytical study of the ASEP with SCPs came with the realization that the matrix product ansatz \cite{Blythe2007,Derrida1993a} can successfully be applied for finding the stationary state. In the seminal paper of Derrida at al.\ \cite{Derrida1993c} the Totally Asymmetric Simple Exclusion Process (TASEP) with second class particles was considered (here $q=0$). Among other important results about microscopic shock measures, it has been shown, that in the limit of zero density of second class particles, they form a weakly bound state. In particular, the distribution of the distance between two second class particles decays as $x^{-\frac{3}{2}}$.

We note, that related questions have been studied in \cite{Chatterjee2008,Rakos2006c} for the Katz Lebowitz Spohn (KLS) model \cite{Katz1984} with probe particles. The KLS model is an exclusion process with nearest neighbour interaction, which, for a special case, reduces to the ASEP, and in this special case the probe particles of \cite{Chatterjee2008,Rakos2006c} are equivalent to the SCPs of the ASEP. Although the probe particles introduced in this model are in general not second class particles, the distance distribution between them was numerically found to decay as $x^{-\frac{3}{2}}$ in \cite{Chatterjee2008}. 

The purpose of this work is to check, whether the $x^{-\frac{3}{2}}$ law is generic for second class particles of systems within the Kardar-Parisi-Zhang (KPZ) university class \cite{Kardar1986}. We consider a one-parameter family of misanthrope processes obtained by a mapping from the KLS model \cite{Evans2005} and study the behaviour of second class particles in this process. In section \ref{sec:model} we introduce the model and define second class particles. In section \ref{sec:stat} the stationary properties (distance distribution of SCPs, density profile seen from a SCP) are studied, while section \ref{sec:dyn} is devoted to dynamical scaling properties. Results are summarized in \ref{sec:sum}. 

\section{Second class particles in general}
\label{sec:SCP}
Second class particles are most often studied in the context of the ASEP. Since other examples for SCPs are less well-known, a general introduction is given here. We consider systems where the occupation number $\eta_i$ of site $i$ is integer, but is not necessarily restricted to be between 0 or 1. For a specific system  $\eta_\mathrm{min}\leq\eta_i\leq\eta_\mathrm{max}$, where $\eta_\mathrm{min}$ and $\eta_\mathrm{max}$ can be any integer, including $\pm\infty$. 

In general, second class particles (SCP) are defined through the basic coupling (described in detail e.g.\ in \cite{Balazs2007}) which works as follows. Two copies of the same system are considered simultaneously which are prepared in an initial state with occupation numbers that differ by one only at a single site. The time evolution of the two copies are then coupled in a way that marginally both of them satisfies the rules of the original model while at every time instance it is ensured that there is only a single site where the occupation number of the two copies differ. For models where this coupling is possible the mismatch site is considered to be occupied by one SCP. 

More than one SCP can be introduced in a similar way by considering two copies that differ at several sites with the condition that the following inequality holds for the occupation numbers 
\begin{equation} \label{major}
 \eta^A_i \geq \eta^B_i.
\end{equation}
Here $\eta^X_i$ denotes the occupation number of copy $X$ at site $i$. The number of SCPs at site $i$ is then defined to be $\eta^A_i - \eta^B_i$. Note that this definition is not restricted to exclusion processes, meaning that $\eta_i$, and therefore also $\eta^A_i - \eta^B_i$, can be any integer number. Due to the conservation law for the original particles, the number of SCPs $\sum_i \eta^A_i - \eta^B_i$ is also conserved. Systems, where property (\ref{major}) at time zero implies that this property remains true for any $t>0$ are called attractive. In non-attractive systems the particle-interpretation of the occupation number differences is not justified, since in this case SCP -- anti-SCP pairs can be created and annihilated (for the attractive case only annihilation is possible). 

Since marginally the two copies evolve according to the original dynamical rules, the presence of second class particles does not influence the dynamics of the original (first class) particles, so they can be considered as passive scalars.  For details about the basic coupling see \cite{Balazs2007}, where a wide family of attractive misanthrope processes are considered with several specific examples.

\section{Model}
\label{sec:model}

\subsection{One-dimensional KLS model}
The model considered here is the one-dimensional Katz-Lebowitz-Spohn (KLS) model \cite{Katz1984}, which is a generalization of the Totally Asymmetric Simple Exclusion Process (TASEP). It is defined on a one-dimensional lattice of length $L$, where each site is either occupied by a particle or empty. During a continuous time evolution particles can hop to the right provided that the target site is empty. In contrast to the TASEP the rate of this jump depends also on the left and right nearest neighbour of the departure and destination sites according to the following rules:
\begin{equation} \label{KLSrules}
\begin{array}{ll}
  0 1 0 0 \rightarrow 0 0 1 0 & \mbox{with rate } 1\\
  1 1 0 1 \rightarrow 1 0 1 1 & \mbox{with rate } 1\\
  0 1 0 1 \rightarrow 0 0 1 1 & \mbox{with rate } 1-\epsilon\\
  1 1 0 0 \rightarrow 1 0 1 0 & \mbox{with rate } 1+\epsilon
\end{array}
\end{equation}
Here 1 and 0 indicates an occupied and an empty site. On top of the on-site exclusion, a positive $\epsilon$ induces a repulsive interaction between particles, while negative epsilon leads to attraction. Note that here the sign of $\epsilon$ is the opposite of what is usually used, i.e., $\epsilon>0$ corresponds to particle repulsion.

For this system additional ``probe'' particles were introduced and the effective interaction between them mediated by the driven fluid (0 and 1) were studied in \cite{Chatterjee2008,Rakos2006c,Levine2005a}. These probes have been defined with the following dynamical rules:
\begin{equation}
\begin{array}{ll}
  1 2 \rightarrow 2 1 & \mbox{with rate } 1\\
  2 0 \rightarrow 0 2 & \mbox{with rate } 1
\end{array}
\end{equation}
where 2 indicates a probe particle. In the special case of $\epsilon = 0$ this model reduces to the TASEP with second class particles, where the second class particles are the probes. However, for any other $\epsilon$ the probes do not behave as second class particles. In particular, the probes do influence the dynamics of 0s and 1s. 

It is relatively easy to see that the model defined in (\ref{KLSrules}) is not attractive (in the sense described in section \ref{sec:SCP}), the above described basic coupling would lead here to a system where SCP--anti-SCP pairs could be created and annihilated. However, as will be described in the next subsection, the model can be mapped to a misanthrope process, which is attractive (for $\epsilon>0$), and where classical SCPs can be introduced (i.e., without SCP creation).

\subsection{Mapping to a misanthrope process}
\label{sec:mapping}
\begin{figure}[h]
\begin{indented}
\item[(a)] 
\scalebox{1} 
{
\begin{pspicture}(0,-0.7892187)(10.09,0.7992188)
\usefont{T1}{ptm}{m}{n}
\rput(8.4114065,0.61078125){$1$}
\usefont{T1}{ptm}{m}{n}
\rput(6.731406,0.61078125){$1$}
\usefont{T1}{ptm}{m}{n}
\rput(4.161406,0.61078125){$1-\epsilon$}
\usefont{T1}{ptm}{m}{n}
\rput(1.6014062,0.5907813){$1+\epsilon$}
\psbezier[linewidth=0.02,arrowsize=0.05291667cm 2.0,arrowlength=1.4,arrowinset=0.4]{->}(1.26,0.06078125)(1.54,0.48078126)(1.82,0.48078126)(2.1,0.06078125)
\psline[linewidth=0.02cm](0.0,-0.77921873)(10.08,-0.77921873)
\psline[linewidth=0.02cm](0.84,-0.63921875)(0.84,-0.77921873)
\psline[linewidth=0.02cm](1.68,-0.63921875)(1.68,-0.77921873)
\psline[linewidth=0.02cm](2.52,-0.63921875)(2.52,-0.77921873)
\psline[linewidth=0.02cm](3.36,-0.63921875)(3.36,-0.77921873)
\psline[linewidth=0.02cm](4.2,-0.63921875)(4.2,-0.77921873)
\psline[linewidth=0.02cm](5.04,-0.63921875)(5.04,-0.77921873)
\psline[linewidth=0.02cm](5.88,-0.63921875)(5.88,-0.77921873)
\psline[linewidth=0.02cm](6.72,-0.63921875)(6.72,-0.77921873)
\psline[linewidth=0.02cm](7.56,-0.63921875)(7.56,-0.77921873)
\psline[linewidth=0.02cm](8.4,-0.63921875)(8.4,-0.77921873)
\psline[linewidth=0.02cm](9.24,-0.63921875)(9.24,-0.77921873)
\psline[linewidth=0.02cm](10.08,-0.63921875)(10.08,-0.77921873)
\pscircle[linewidth=0.02,dimen=outer,fillstyle=solid,fillcolor=black](0.42,-0.35921875){0.29}
\pscircle[linewidth=0.02,dimen=outer,fillstyle=solid,fillcolor=black](1.26,-0.35921875){0.29}
\pscircle[linewidth=0.02,dimen=outer,fillstyle=solid,fillcolor=black](3.78,-0.35921875){0.29}
\pscircle[linewidth=0.02,dimen=outer,fillstyle=solid,fillcolor=black](5.46,-0.35921875){0.29}
\pscircle[linewidth=0.02,dimen=outer,fillstyle=solid,fillcolor=black](6.3,-0.35921875){0.29}
\pscircle[linewidth=0.02,dimen=outer,fillstyle=solid,fillcolor=black](7.98,-0.35921875){0.29}
\psbezier[linewidth=0.02,arrowsize=0.05291667cm 2.0,arrowlength=1.4,arrowinset=0.4]{->}(3.78,0.06078125)(4.06,0.48078126)(4.34,0.48078126)(4.62,0.06078125)
\psbezier[linewidth=0.02,arrowsize=0.05291667cm 2.0,arrowlength=1.4,arrowinset=0.4]{->}(6.3,0.06078125)(6.58,0.48078126)(6.86,0.48078126)(7.14,0.06078125)
\psbezier[linewidth=0.02,arrowsize=0.05291667cm 2.0,arrowlength=1.4,arrowinset=0.4]{->}(7.98,0.06078125)(8.26,0.48078126)(8.54,0.48078126)(8.82,0.06078125)
\psline[linewidth=0.02cm](0.0,-0.63921875)(0.0,-0.77921873)
\end{pspicture} 
}
\item[]
\item[(b)] 
\scalebox{1} 
{
\begin{pspicture}(0,-1.3392187)(5.1709375,1.3692187)
\psline[linewidth=0.02cm](0.1209375,-1.3292187)(5.1609373,-1.3292187)
\psline[linewidth=0.02cm](0.1209375,-1.1892188)(0.1209375,-1.3292187)
\psline[linewidth=0.02cm](0.9609375,-1.1892188)(0.9609375,-1.3292187)
\psline[linewidth=0.02cm](1.8009375,-1.1892188)(1.8009375,-1.3292187)
\psline[linewidth=0.02cm](2.6409376,-1.1892188)(2.6409376,-1.3292187)
\psline[linewidth=0.02cm](3.4809375,-1.1892188)(3.4809375,-1.3292187)
\psline[linewidth=0.02cm](4.3209376,-1.1892188)(4.3209376,-1.3292187)
\psline[linewidth=0.02cm](5.1609373,-1.1892188)(5.1609373,-1.3292187)
\pscircle[linewidth=0.04,dimen=outer,fillstyle=solid,fillcolor=black](0.5409375,-0.9092187){0.3}
\pscircle[linewidth=0.04,dimen=outer,fillstyle=solid,fillcolor=black](0.5409375,0.21078125){0.3}
\pscircle[linewidth=0.04,dimen=outer,fillstyle=solid,fillcolor=black](2.2209375,-0.9092187){0.3}
\pscircle[linewidth=0.04,dimen=outer,fillstyle=solid,fillcolor=black](3.0609374,-0.9092187){0.3}
\pscircle[linewidth=0.04,dimen=outer,fillstyle=solid,fillcolor=black](3.0609374,-0.20921876){0.3}
\pscircle[linewidth=0.04,dimen=outer,fillstyle=solid,fillcolor=black](3.9009376,-0.9092187){0.3}
\psbezier[linewidth=0.02,arrowsize=0.05291667cm 2.0,arrowlength=1.4,arrowinset=0.4]{->}(0.5409375,0.63078123)(0.8209375,1.0507812)(1.1009375,1.0507812)(1.3809375,0.63078123)
\psbezier[linewidth=0.02,arrowsize=0.05291667cm 2.0,arrowlength=1.4,arrowinset=0.4]{->}(2.2209375,0.21078125)(2.5009375,0.63078123)(2.7809374,0.63078123)(3.0609374,0.21078125)
\psbezier[linewidth=0.02,arrowsize=0.05291667cm 2.0,arrowlength=1.4,arrowinset=0.4]{->}(3.0609374,0.21078125)(3.3409376,0.63078123)(3.6209376,0.63078123)(3.9009376,0.21078125)
\psbezier[linewidth=0.02,arrowsize=0.05291667cm 2.0,arrowlength=1.4,arrowinset=0.4]{->}(3.9009376,-0.48921874)(4.1809373,-0.06921875)(4.4609375,-0.06921875)(4.7409377,-0.48921874)
\usefont{T1}{ptm}{m}{n}
\rput(4.3323436,0.06078125){$1$}
\usefont{T1}{ptm}{m}{n}
\rput(0.88234377,1.1807812){$1+\epsilon$}
\usefont{T1}{ptm}{m}{n}
\rput(2.6023438,0.74078125){$1-\epsilon$}
\usefont{T1}{ptm}{m}{n}
\rput(3.5123436,0.7607812){$1$}
\psdots[dotsize=0.08](0.5409375,-0.20921876)
\psdots[dotsize=0.08](0.5409375,-0.34921876)
\psdots[dotsize=0.08](0.5409375,-0.48921874)
\end{pspicture} 
}
\end{indented}
\caption{(a) Transition rates of the KLS model. (b) Transition rates of the corresponding misanthrope process. The rates depend on the occupation numbers of both the departure and the destination sites. \label{fig:mapping}}
\end{figure}
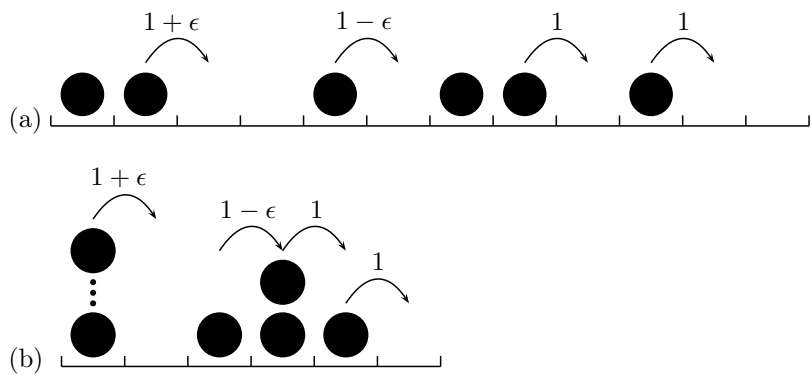

The mapping, which has been used also in \cite{Evans2005} consists of the following steps:
\begin{itemize}
 \item The vacancies of model (\ref{KLSrules}) (i.e., sites with occupation number zero) are mapped to the lattice sites of the new model.
 \item The length of the uninterrupted sequence of occupied sites to the right of a vacancy in the old model is considered to be the occupation number of the corresponding site in the new model.
\end{itemize}
For an illustration see figure~\ref{fig:mapping}. As a result we obtain a model where the occupation number $\eta_i$ can be any nonnegative integer ($\eta_\mathrm{min}=0$, $\eta_\mathrm{max}=\infty$) with the following dynamical rules:
\begin{equation}
 \begin{array}{llll}
   \eta_i,0 &\rightarrow \eta_i-1,1 &\mbox{with rate } 1+\epsilon & \mbox{if } \eta_i\geq 2 \\ 
   \eta_i,\eta_{i+1} &\rightarrow \eta_i-1,\eta_{i+1}+1 &\mbox{with rate } 1 & \mbox{if } \eta_i\geq 2 \mbox{ and } \eta_{i+1}\geq 1\\ 
   1,\eta_{i+1} &\rightarrow 0,\eta_{i+1}+1 &\mbox{with rate } 1-\epsilon & \mbox{if } \eta_{i+1}\geq 1\\ 
   1,0 &\rightarrow 0,1 &\mbox{with rate } 1 
 \end{array}
\end{equation}
This defines a misanthrope process where the hopping rate of a particle is a function of the occupation number of the departure and destination sites. It can be seen that this model, in contrast to the KLS model (\ref{KLSrules}), is attractive for $\epsilon>0$. (Note that here attractivity is meant in the sense described in section \ref{sec:SCP}, the particles themselves interact repulsively for $\epsilon>0$.) The main property that provides attractivity is that the transition rate $r(\eta_i,\eta_{i+1})$, which is the rate from configuration $\eta_i,\eta_{i+1}$ to configuration $\eta_i-1,\eta_{i+1}+1$ is a nondecreasing function of $\eta_i$ and a non-increasing function of $\eta_{i+1}$. For details see \cite{Balazs2007}. Notice that the particle-hole symmetry of the original exclusion process is lost by the mapping. 

\section{Stationary properties}
\label{sec:stat}

In this section we consider two SCPs in the system with periodic boundaries and the distance between these two particles is studied. It is known that the dynamics of second class particles is closely related to density and current fluctuations in a system of the KPZ class \cite{Balazs2007}, therefore results for SCPs reflect the nature of these fluctuations. In \cite{Derrida1993c} it was shown analytically that in the ASEP the (non-cumulative) probability distribution of the distance between two second class particles in a stationary situation has an algebraic decay for large distances with a power $3/2$:
\begin{equation} \label{universal_distance_distribution}
 {\cal P}(x) \sim x^{-\frac{3}{2}}
\end{equation}
(Here it is assumed that first the size $L$ of the periodic system is taken to infinity, so $1\ll x\ll L$.)

The question we study is whether (\ref{universal_distance_distribution}) remains true in a more general case. For this reason we study the misanthrope process introduced in section \ref{sec:model} with $\epsilon>0$. Most studies of this kind assume half-filling for the exclusion particles which maps to density $\sigma=1$ in the misanthrope process. We restrict our study for this special case. 

\begin{figure}
\begin{indented}
 \item[]  \input{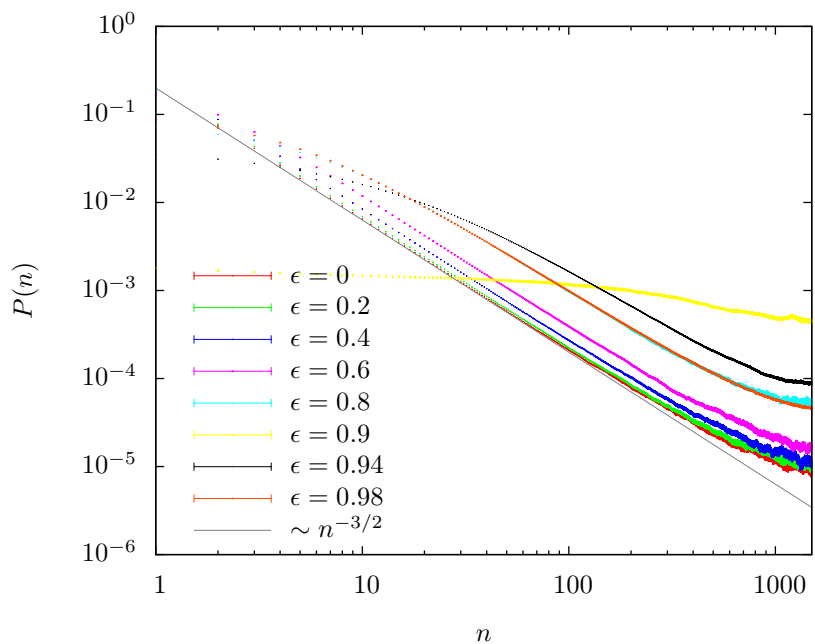}
\end{indented}
\caption{\label{fig:stationary_distribution} Shown is the measured distribution of the distance between two second class particles for various values of $\epsilon$ in a periodic system of 3000 sites with $\sigma=1$. The result suggests that the exponent of the algebraic decay is universal (-3/2). Increasing the value of $\epsilon$ the current-density relation becomes convex at $\rho=1$. At this transition point the universality is not expected to hold}
\end{figure}

In order to check the above property we did numerical simulations in a periodic system with $N$ sites. Initially the system has been prepared in a state where the number of particles is $N$ with two extra SCPs (this corresponds to half filling in the exclusion model). Technically this means that we consider two copies in parallel: copy ``A'' holds $N$ particles, while copy ``B'' is a clone of A with two extra particles, as described in section \ref{sec:SCP}. Initially the particles are placed uniformly in the system, and after a sufficient relaxation time the distance distribution between the two SCPs is measured. Results are shown in figure \ref{fig:stationary_distribution}.

\begin{figure}
\input{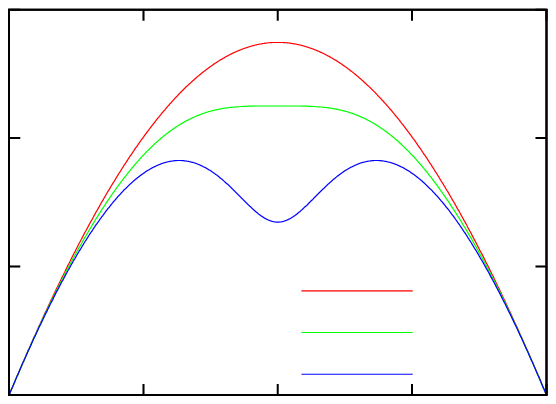} \input{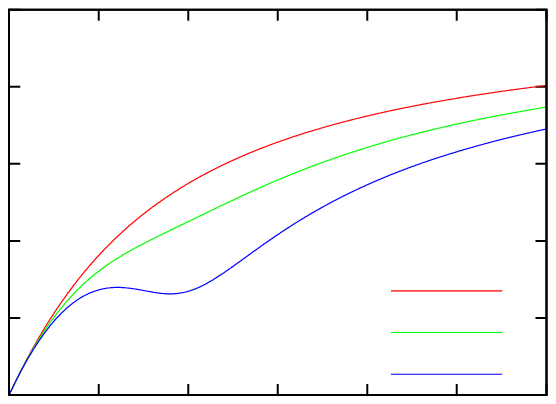}
\caption{\label{fig:currents} Shown is the current--density relation of the KLS model (a) and the corresponding misanthrope process (b). The functions $j(\rho)$ and $J(\sigma)$ are concave for $\epsilon<0.8$. For half filling ($\rho=0.5$, $\sigma=1$) the second derivative of $j(\rho)$ and $J(\sigma)$  becomes zero  at $\epsilon=0.8$, and further increasing $\epsilon$ it becomes positive.}
\end{figure}

For most values of $\epsilon$ it can be seen that the distance distribution follows (\ref{universal_distance_distribution}), which suggests that this value is universal within the KPZ class. However, for a specific value of $\epsilon$ (which appears to be near 0.9) a different behaviour is observed, which can be understood by looking at the stationary current--density relation. For systems of the KPZ class this relation is strictly nonlinear. The current--density relation $j(\rho)$ for the KPZ model, and $J(\sigma)$ for the related misanthrope process is shown in figure \ref{fig:currents}. The mapping  relates the two currents and densities as follows:
\begin{equation}
  \sigma=\frac{\rho}{1-\rho} \qquad J=\frac{j}{1-\rho},
\end{equation}
where $j$ and $\rho$ are the current and the density in the KLS model while in the misanthrope process these are denoted by $J$ and $\sigma$ correspondingly. An interesting property of this mapping is that the sign of the second derivative of $j(\rho)$ always agrees with that of the second derivative of $J(\sigma)$ at the corresponding value of $\sigma$.

The analytical form of $j(\rho)$ and therefore also of $J(\sigma)$ is known (see \cite{Hager2001}). It can be shown that the current-density relation is concave for $\epsilon<0.8$, while for $0.8<\epsilon<1$ the second derivative becomes positive at $\rho=0.5$ ($\sigma=1$). Near $\epsilon=0.8$ and $\sigma=1$ the second derivative vanishes which suggests a crossover to the Edwards-Wilkinson universality class, and this explains that the behaviour of ${\cal P}(n)$ is different at a special value of $\epsilon$ (in the Edwards-Wilkinson class one expects no effective interaction between SCPs, which is consistent with the results of figure \ref{fig:stationary_distribution}). However, this value appears to be about $0.9$ rather than $0.8$, as suggested by the theory. It would be interesting to find a satisfactory explanation for this discrepancy.

\begin{figure}
\begin{indented}
 \item[]  \input{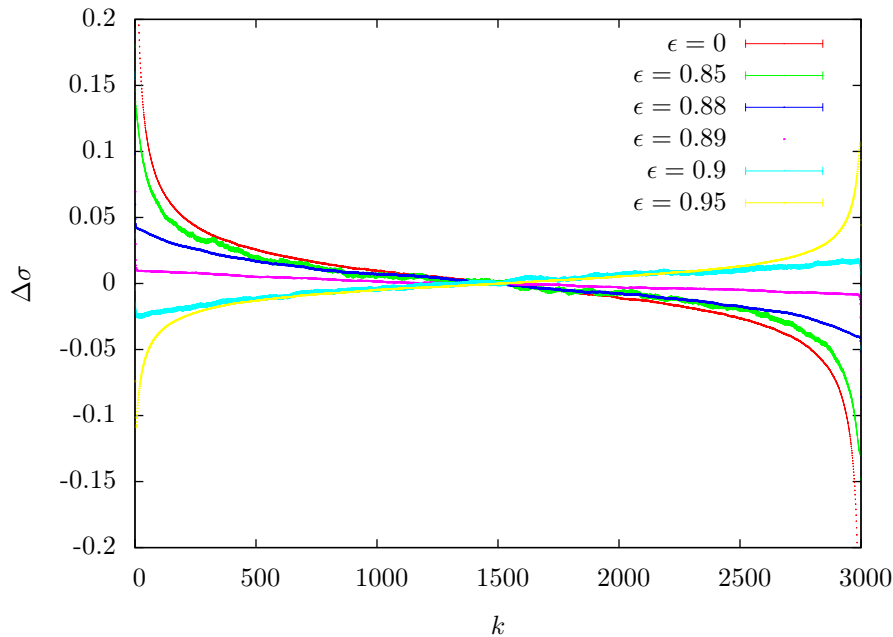}
\end{indented}
\caption{\label{fig:profiles} The density profile as seen from a single SCP was measured in a periodic system of 3000 sites with $\sigma=1$ and several values of $\epsilon$ ($\Delta \sigma=\sigma-1$). At a specific value $\epsilon_\mathrm{c}$ the profile becomes flat which indicates a crossover to the Edwards-Wilkinson universality class.}
\end{figure}

SCPs in general are attracted by high density gradients and the effective interaction between SCPs is mediated by the density fluctuations of the original particles. This can be clearly seen by measuring the density seen from a single SCP. Such density profiles are shown in figure \ref{fig:profiles}. Since the position of the SCP is highly correlated with density fluctuations, the profile measured from the SCP is in general far from being flat (although the stationary density profile, as measured by a standing observer is flat). The average speed of the SCP in a homogeneous environment with density $\sigma$ is given by $v(\sigma)=J'(\sigma)$, which is also called the characteristic speed. At the point where the second derivative $J''(\sigma)$ becomes zero the SCP becomes insensitive to (small) density fluctuations therefore one expects that the density profile seen from this particle becomes flat. This is what we see in figure \ref{fig:profiles}. Here the crossover point appears to be the same ($\epsilon_\mathrm{c}\approx 0.9$) as in figure \ref{fig:stationary_distribution} although the theory predicts $\epsilon_\mathrm{c}=0.8$.

Notice that even though for $\epsilon=0$ the KLS model reduces to the TASEP, the SCP of the corresponding misanthrope process is not equivalent to the SCP of the TASEP.   

\section{Dynamical properties}
\label{sec:dyn}

It is interesting to see how the distance between two SCPs changes in time if they start at the same position. Similar studies for other sort of probe particles were done in \cite{Rakos2006c} and \cite{Chatterjee2008}. We consider again the distance distribution between two SCPs but now as a function of time. In the simulation the system is prepared in a random initial configuration with $\sigma=1$ and two SCPs, just like in the previous section. After a sufficiently long relaxation time the two SCPs are monitored and the time is set to zero when the two particles occupy the same site (this initialization procedure agrees with the one used in \cite{Chatterjee2008}). The distance between the SCPs is measured and the distribution is calculated for sufficiently large number of independent realizations. 

\begin{figure}
\begin{indented}
 \item[]  \input{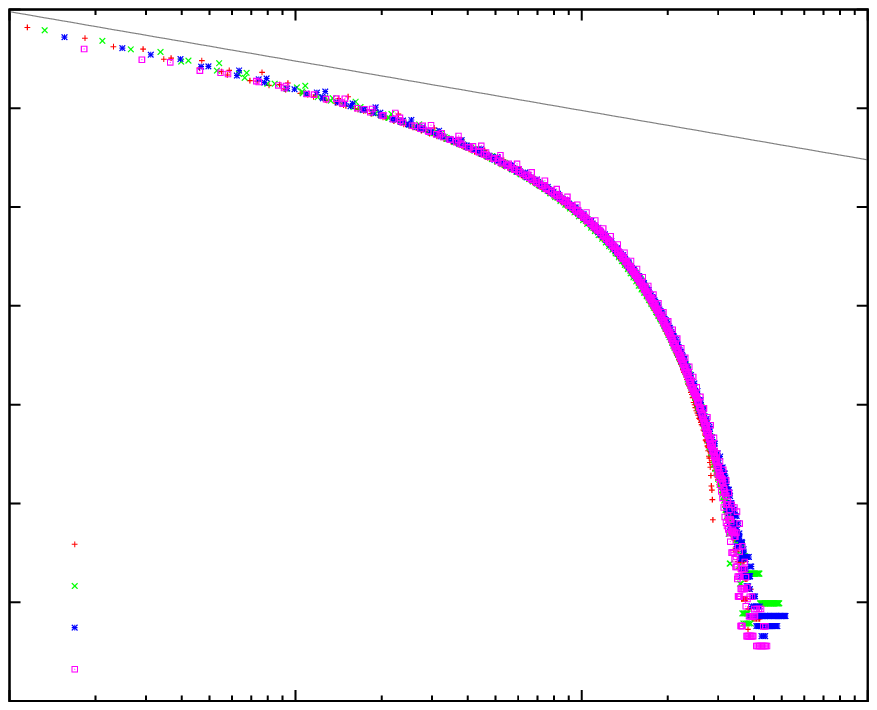}
\end{indented}
\caption{Shown is the scaling collapse of the cumulative distribution function of the distance between two second class particles for various values of $t$ and $\epsilon$. \label{fig:scaling}}
\end{figure}

Our scaling ansatz for ${\cal P}(x,t)$ for a specific $\epsilon$ is
\begin{equation}
 {\cal P}(x,t)= F\left(\frac{x}{t^{1/z}}\right) t^{-\frac{3}{2z}}, 
\end{equation}
where $z$ is the dynamical critical exponent and $F$ is a scaling function. The scaling function $F(y)$ for small arguments $y$ is assumed to behave as $F(y)\sim y^{-\frac{3}{2}}$ in accordance with the stationary behaviour. For the cumulative distribution function $\tilde{\cal P}(n,t)=\sum_{\ell=n}^\infty {\cal P}(\ell,t)$ the above ansatz implies
\begin{equation}
 \tilde{\cal P}(n,t)= G\left(\frac{n}{t^{1/z}}\right) t^{-\frac{1}{2z}}
\end{equation}
with another scaling function $G$, for which $G(y)\sim y^{-\frac{1}{2}}$ if $y\ll 1$. This scaling ansatz has been tested and confirmed in simulations. Furthermore, it is shown that not only the exponents but also the scaling function $G$ is universal (up to non-universal scaling amplitudes $a(\epsilon)$ and $b(\epsilon)$). Results are shown in figure \ref{fig:scaling}, where $b(\epsilon)t^{\frac{1}{2z}}\tilde{\cal P}(n,t)$ is plotted against $\frac{n}{t^{1/z}}$ with the dynamical critical exponent of the KPZ class $z=3/2$. The amplitudes $a(\epsilon)$ and $b(\epsilon)$ were fitted by hand for each value of $\epsilon$. We note that the scaling collapse is less convincing for larger values of $\epsilon$ ($\epsilon \gtrsim 0.7$) which might be due to the crossover expected at the point where the current-density relation turns from concave to convex. Close to this critical point the convergence is expected to be slow.

\section{Summary}
\label{sec:sum}

We considered a one-parameter family of asymmetric misanthrope processes, which are derived from the one-dimensional Katz-Lebowitz-Spohn (KLS) model \cite{Katz1984}. Although the standard KLS model is not attractive in the sense, that second class particles cannot be introduced with conservation, using a mapping to a misanthrope process, the resulting model (in a limited range of parameters) is attractive, thus second class particles can be defined using the basic coupling. 

We studied numerically the effective interaction between second class particles of this misanthrope process. It has been fond that the distance distribution between two second class particles has a power law tail with an exponent $-\frac{3}{2}$. This exponent has already been found to be exact in the Asymmetric Simple Exclusion Process \cite{Derrida1993c}. Note that although in a special case the KLS model reduces to the ASEP, the second class particles of the misanthrope process are not equivalent to those of the exclusion process. It is interesting to compare this result with a related one for probe particles in the one-dimensional KLS model \cite{Chatterjee2008}, where similar distance distribution is found for probe particles, although these are not second class particles. 

In addition, the dynamical scaling function of the distance distribution has been obtained numerically. We found a good data collapse for several values of a model parameter, which suggests that this scaling function is universal within the KPZ universality class. 

For a special value of the model parameter $\epsilon$, the quadratic non-linearity in the (exactly known) current-density relation vanishes (at $\sigma=1$), which suggests a crossover to another universality class (Edwards-Wilkinson). The crossover is observed in the simulation results, however, the location of this crossover shows a slight mismatch with the theory. This discrepancy could be due to finite size effects. 

The study of this paper could easily be extended to different densities of first class particles, and/or partially asymmetric dynamics. The behaviour for these more general cases is expected to be similar to what we found for the considered spacial case. Another, more interesting generalization would be to consider more second class particles and study the crossover between finite number and finite density of second class particles. 

\section{Acknowledgements}
I would like to thank Gunter Sch\"utz, M\'arton Bal\'azs, and Sakuntala Chatterjee for fruitful and stimulating discussions. 
Financial support of the Hungarian Research Fund (OTKA) under grant number \mbox{PD-72607} and \mbox{PD-78433} is greatly acknowledged as well as support from the Bolyai Fellowship. 
\section*{References}
\bibliographystyle{iopart-num}
\bibliography{SCP}

\end{document}